\documentclass[10pt]{wlscirep}
\usepackage[utf8]{inputenc}
\usepackage[T1]{fontenc}

\title{Sound Reconstruction via Optical Multi-Mode Fiber}

\author[1,+]{Ege Küçükkömürcü}
\author[2,*]{Berk Nezir Gün}
\author[2]{Emre Yüce}

\affil{Acoustics MSc., École Centrale de Lyon, 69130, Écully, France}
\affil[2]{Programmable Photonics Group, Department of Physics, Middle East Technical University, 06800, Ankara, Turkey}

\affil[+]{ege.kucukkomurcu@etu.ec-lyon.fr}
\affil[*]{gun@metu.edu.tr}

\begin{abstract}

Sound reconstruction via arbitrary objects has been a popular method in recent years, based on the recording of scattered light from the target object with a high-speed detector. In this work, we demonstrate the use of multi-mode fiber as a medium that enables reconstruction at a much further distance. By placing a speaker near the fiber and using a high-speed camera with a maximum region of interest, we show that it is possible to reconstruct the sound without relying on specific grain information in the speckle. Basic filtration techniques and spectral subtraction methods are employed to improve the signal quality. After applying band-stop and high-pass filters, the magnitude-squared coherence between the original sound and captured data is 0.8 at fundamental frequencies. The proof-of-concept method that we introduce here can be extended to monitor the vibrations of earth's crust and civilian buildings for structural safety.

\end{abstract}
\begin{document}

\flushbottom
\maketitle
\thispagestyle{empty}
\section*{Introduction} Sound is a pressure wave that excites all translational, rotational, and vibrational modes of atoms in a medium. What we hear is the motion of the atoms, governed by the vibrational modes. Similar to an eardrum, a microphone also has a diaphragm that is sensitive to vibrations across a wide range of frequencies. The diaphragm of the microphone is excited by the vibrations and converted into an electrical signal via an inductive coil. In fact, the motion of the diaphragm can also be visualized and converted into sound, enabling sound reconstruction from consecutive images \cite{Davis2014VisualMic}. Capturing vibrations from arbitrary shaped objects might be challenging; however, there are many advantages and applications.

Sound reconstruction through optical means is appealing due to its wider range of detection than that of classical microphones and its durability in harsh environmental conditions. An additional plus for optical techniques is that the optical system carries information about the sound source's location, contrary to a classical microphone. With optical methods, real-time sound reconstruction might also be possible, depending on the complexity of data processing and computational power. The advantages of optical techniques to get vibrations have catalyzed various applications in non-destructive testing\cite{civil}, medical diagnosis\cite{photoacoustic}, optical coherence tomography\cite{OCT}, and security\cite{lamphone}.

In our approach to reconstructing sound optically, we utilize Laser Doppler Vibrometry (LDV) principles \cite{ldv}, specifically with a focus on the inherent sensitivity of multi-mode fiber. Unlike traditional LDV methods that involve complex setups \cite{fiberldv1,fiberldv2}, our method provides a simpler alternative for sound reconstruction without the need for a second light path. By leveraging the millimeter range between the source and the multi-mode fiber, our approach captures and reconstructs sound using only a high-speed camera, multi-mode fiber, and light source, making it more straightforward and efficient in capturing the vibration characteristics of the object. This emphasis on simplicity and effectiveness of our method as a viable and practical solution for optical sound reconstruction.

Various signal processing methods are proposed to properly reconstruct sound from surface vibrations. Some options are phase-based algorithms\cite{9453154}\cite{9551943}, singular value decomposition\cite{svd}, digital image correlation\cite{imcor}, and machine learning algorithms\cite{9549233}. Within our scenario, our inclination lies in computing the average of pixel values encompassing the captured speckle pattern as a preliminary step for data processing prior to embarking on the sound reconstruction endeavor. We use an 8-bit camera that provides grayscale images with a dynamic range between 0-255.

\section*{Experimental Setup}
We use a telecom laser at L=1550 nm as the light source. The speaker, whose maximum power is 90 dBa, is placed such that it makes a 2mm distance from the multi-mode fiber (Thorlabs FG105LCA). The fiber vibrates as the speaker plays the sound, and so does the speckle pattern within. To detect the frequencies that the sound wave produces through fiber, we collect the frames taken by a high-speed infrared camera (Goldeye-008) at the output. We use a computer with an Intel Core i7-7700 processor and 16 GB RAM for postprocessing.

\section*{Method}
Our primary objective is to capture high-frequency components with maximum fidelity to accurately reconstruct the emitted sound. To achieve this, we operate the camera at its highest frame rate, directing our attention to a localized region that harbors sufficient speckle information for the subsequent reconstruction process. Having maximum fps requires the camera resolution to be set to a subregion of 128x8 pixels with 500{$\mu$}s exposure time. To pick a proper subregion, we utilize full-resolution snapshots taken at different times when the sound is on. We obtain the subregion that we will use moving forward by consecutively taking the differences of these 8-bit snapshots for every 128x8 pixels within the larger 320x256 pixel frame and selecting the subregion with the highest maximum intensity value.

\begin{figure}[ht]
\centering
\includegraphics[width=0.9\textwidth]{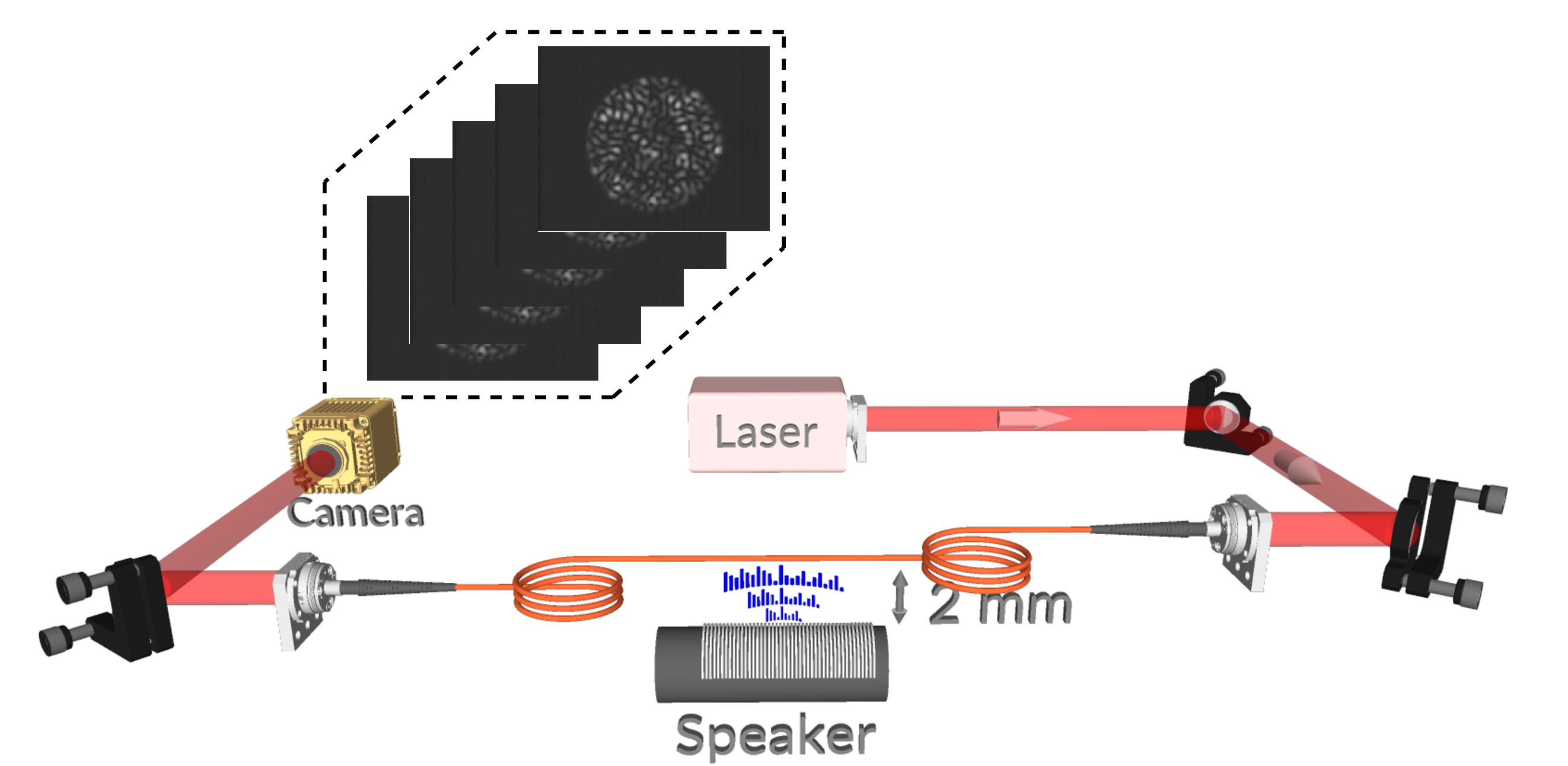}
\caption{The experimental setup used for the reconstruction of sound through capturing silent video of the vibrating laser beam. The infrared light beam is emitted from the laser source and guided towards the coupler (indicated by the arrow). The surface vibrations originating from speaker sound are transmitted to the high-speed camera with the same optical principles. The intensity values of speckle patterns captured by the camera are multiplied by 5 for enhanced visibility.}
\label{fig:setup}
\end{figure}

We play the 5-second piano version of Inspector Gadget's Theme near the multi-mode fiber. To address the absence of a trigger option for synchronizing the timing of the original sound and the recording within our system, a post-recording cross-correlation method is employed. When initiating the recording process, we did not have precise knowledge of which segment of the signal was being captured. Therefore, after the recording, cross-correlation is utilized to align the captured vibrations with the original sound data. To facilitate this alignment, the original audio signal undergoes temporal downsampling from 41.8 kHz to 1.9 kHz, ensuring compatibility with the sampling rate of our recording system.

%after the data is taken, we deploy the cross-correlation method to align the data of captured vibrations to the original sound data. 

\begin{equation}
\widehat{R}_{xy}(m)=
\begin{cases}
\displaystyle\sum_{n = 0}^{N - m - 1}x_{n+m}y_{n}^{*} & \text{if } m\ge 0, \\
\vphantom{\displaystyle\sum_{n = 0}^{N - m - 1}}\widehat{R}_{xy}^{*}(-m) & \text{if } m < 0,
\end{cases}
\label{eq1}
\end{equation}

where ${R}_{xy}(m)$ is the cross-correlation sequence at lag m, $x_{n}$ and $y_{n}$ are the original and recorded data being correlated. We apply m amount of shift that gives us the maximum value of cross-correlation for the alignment. 

\begin{figure*}[ht]
\centering
\setlength\fboxsep{0pt}
\setlength\fboxrule{0pt}
\fbox{\includegraphics[width=0.9\linewidth]{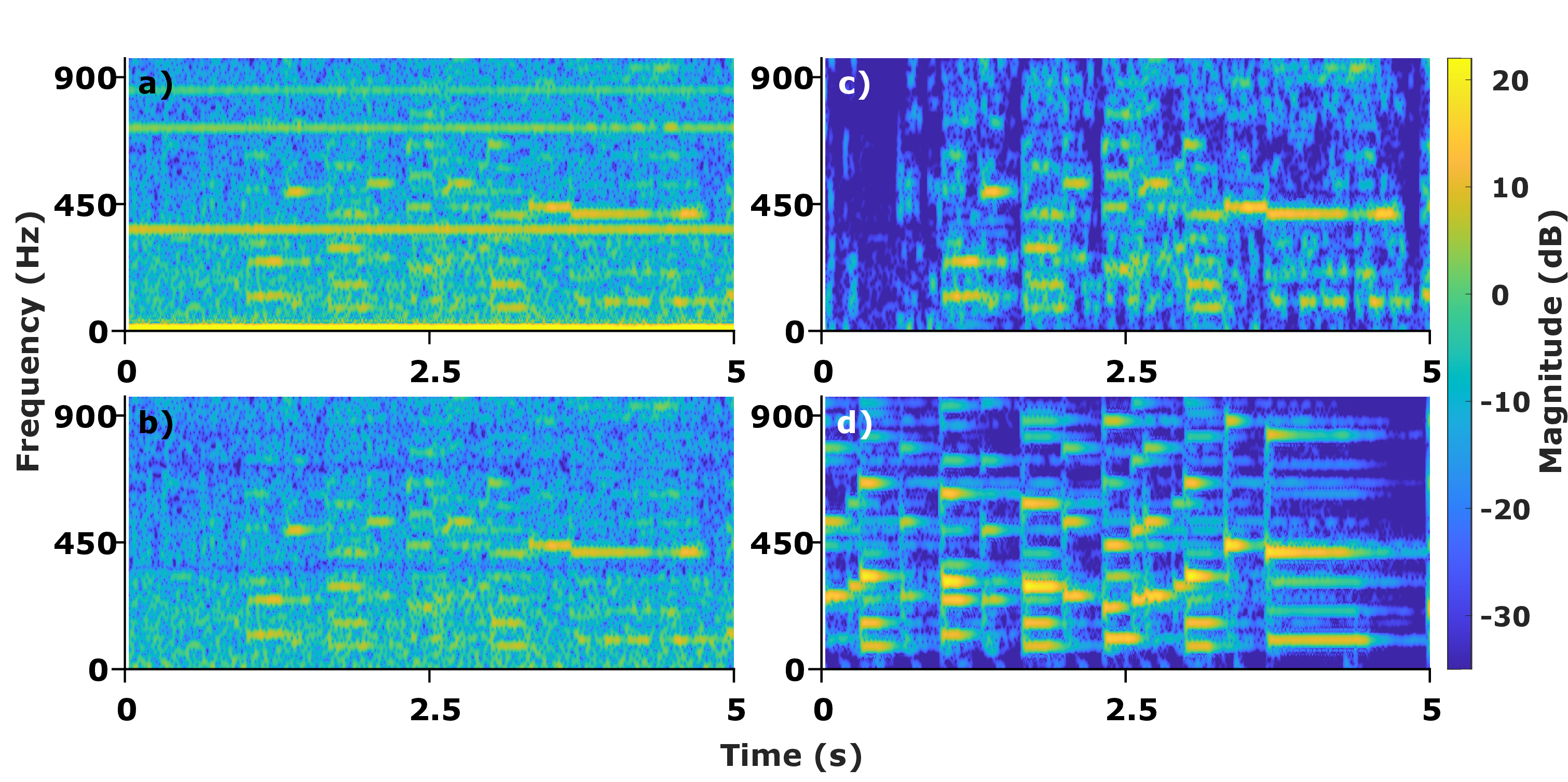}}
\caption{Time(s) vs. frequency(Hz) graphs of (a) captured data without filter, (b) data after broadband and band-stop filters are applied, (c) data after spectral subtraction method is used in addition to the filters applied in (b). (d) Time(s) vs. frequency(Hz) graph of original sound data played by the speaker.}
\label{fig:result}
\end{figure*}

In our setup, we use a high-speed camera to capture the speckle pattern, which is the interference pattern produced by the laser beam reflecting off the vibrating surface. To ensure that we capture enough information and do not average out the data, we confirm that the size of the grains in the speckle is larger than the single-pixel size of the camera. We capture a total of 10.000 frames of speckle patterns during the play of the 5-second record. After the frames are captured from the designated area of 128x8 pixels, a comprehensive set of filtering techniques is applied to acquire optimal results. These techniques, commonly used in signal processing, are specifically chosen to address the presence of noise related to pulses, as evidenced in Figure \ref{fig:result}(a). The frequencies of these pulses are 360 Hz, 720 Hz (2nd harmonic of 360 Hz), and 840 Hz. We also reconstructed the signal itself before playing any sound near the multi-mode fiber. Observing the signal in Fourier domain we saw that we have setup related noises in these certain frequencies.

\begin{figure}[ht]
\centering
\includegraphics[width=0.9\textwidth]{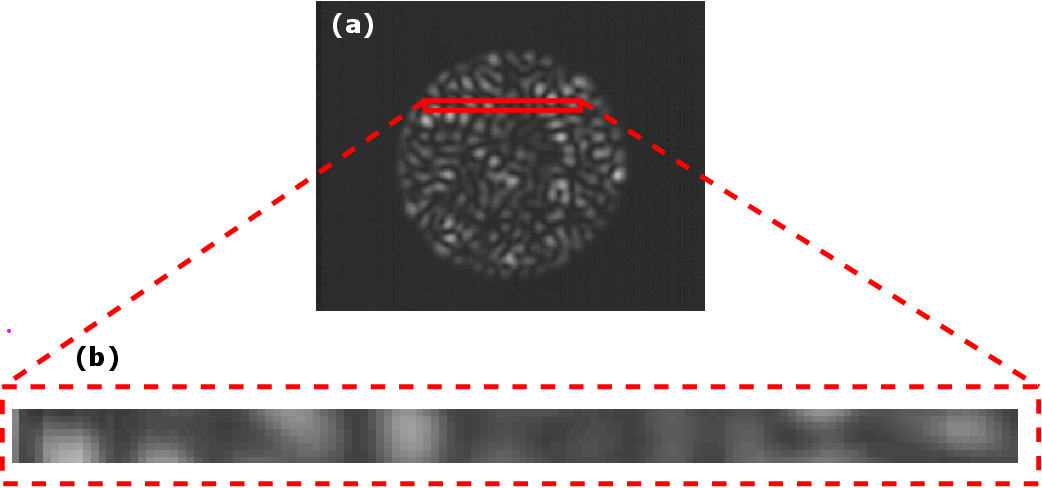}
\caption{(a) 320x256 pixel speckle image with 5x intensity scaling. (b) 128x8 pixel speckle image used to gather data from the region shown.}
\label{fig:speckle}
\end{figure}

\begin{itemize}
\item{Band-stop and high-pass filtering}
\end{itemize}

A band-stop filtering technique was implemented to suppress the noise in the system. This method removes specific frequency ranges, in this case, 360 Hz, 720 Hz, and 840 Hz, to effectively eliminate the noise. A high-pass filter was also applied to remove excessive noise at low frequencies, specifically those below 30 Hz, as seen in Figure \ref{fig:result}(a). Using band-stop and high-pass filtering allowed noise reduction, thus enhancing signal quality.
\begin{itemize}
\item{Audio enhancement}
\end{itemize}

Following the implementation of the high-pass and band-stop filters, the signal was normalized between -1 and 1 to optimize its volume level. This step is essential as normalization ensures the signal is within a consistent range, preventing a possible distortion. To enhance sound quality further, we employed the spectral subtraction method \cite{Bollpaper,Bollfile}, a technique used to remove background noise from sound signals by estimating the background noise's power spectral density (PSD) and subtracting it from the total PSD of the signal.

\section*{Results}

Figure \ref{fig:result}(a) shows the short-time Fourier transform (STFT) of captured vibrations on the fiber. Those vibrations are caused by the external sound we introduced and the electronic devices on the optical table as a noise factor. As seen, below 30 Hz, huge noise was detected in addition to relatively less but consistent noise over time around 360 Hz, 720 Hz, and 840 Hz. Figure \ref{fig:result}(b) shows the STFT of captured data after filtering constant noise frequencies. Here, the magnitudes of frequencies are not apparent compared to Figure \ref{fig:result}(d), which represents the original data. To quantify the similarity between the original and captured data, listening to the reconstructed sound one can perceive the song that is being played. To justify our results, as a metric we used magnitude-squared coherence. The magnitude-squared coherence is a function of the power spectral densities, $P_{xx}(f)$ and $P_{yy}(f)$, and the cross power spectral density, $P_{xy}(f)$, of x (original sound) and y (reconstructed sound):

\begin{equation}
C_{xy}(f) = \frac{\lvert P_{xy}(f) \rvert^2}{P_{xx}(f)P_{yy}(f)}
\end{equation}

Within our study, we examine several critical aspects of our system's performance. The analysis of the region of interest reveals that among the potential 255 grey levels, the maximum grey level stands at 35, while the mean value settles at 17. However, the accuracy of our system is challenged by the presence of environmental noise, necessitating the application of appropriate noise reduction strategies. To address this, we employ a combination of bandstop and high-pass filters, effectively mitigating specific noise components.

\begin{figure*}[ht]
\centering
\setlength\fboxsep{0pt}
\setlength\fboxrule{0pt}
\fbox{\includegraphics[width=0.9\linewidth]{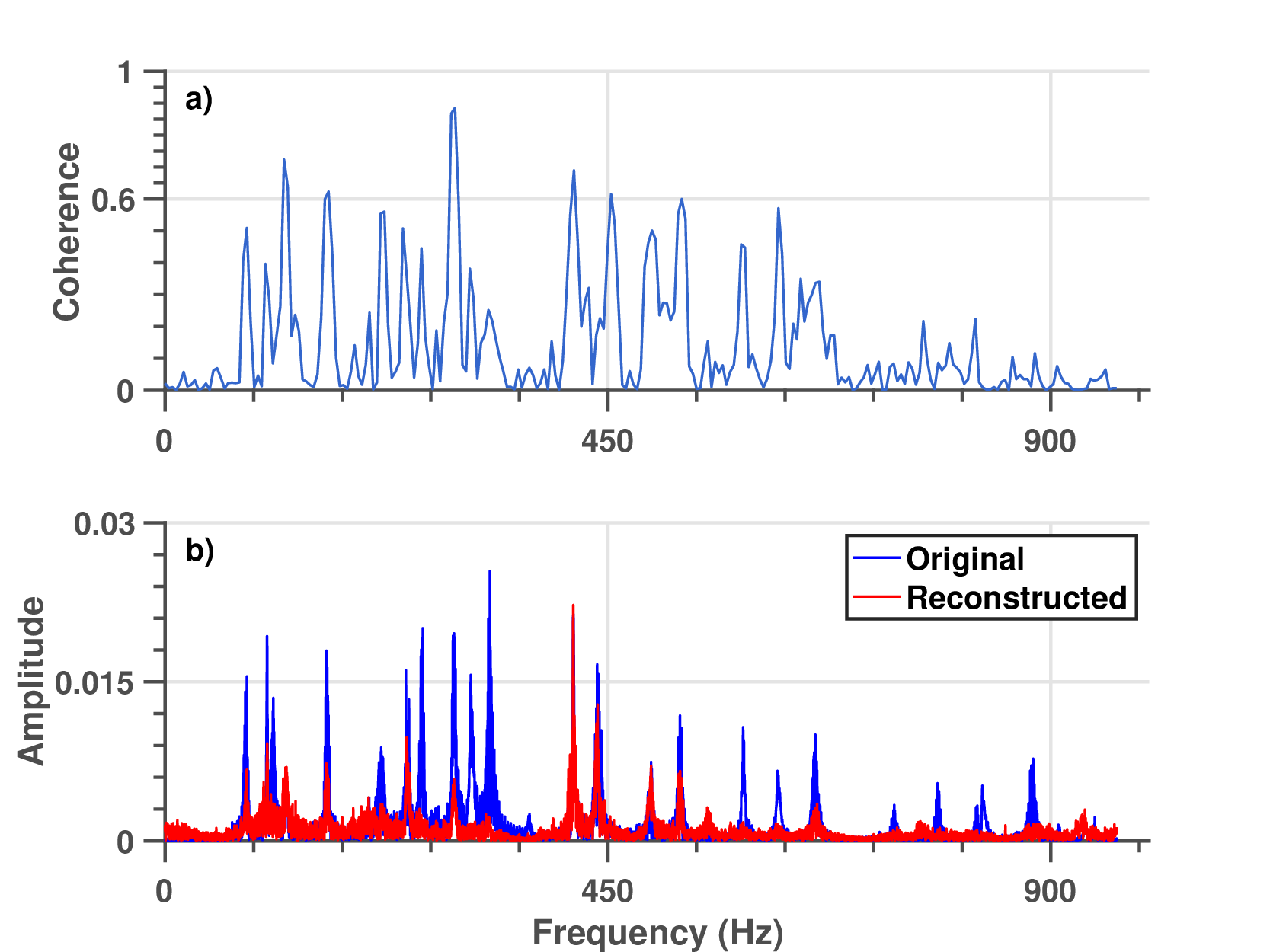}}
\caption{Analysis of Reconstructed and Original Signals (a) magnitude-squared coherence between signals over frequencies (b) Signals in Fourier domain, blue: original signal, red: reconstructed signal.}
\label{fig:coherence}
\end{figure*}

Investigating the magnitude-squared coherence between the two signals reveals noteworthy insights. The coherence is close to 1 at main frequencies of the sound, which are essential for perceiving the played song. For instance, examining the fundamental frequencies of the original sound in Figure \ref{fig:coherence}(a), at 80 Hz, we observe a coherence of 0.5, at 100 Hz a coherence of 0.7, and at 300 Hz a coherence of 0.8. This also coincides with the Figure \ref{fig:result}(c) and Figure \ref{fig:result}(d). 

As monitored in Figure \ref{fig:result}, we observe that the frequencies between 350 Hz and 550 Hz are reconstructed with higher magnitudes than other frequencies. Based on our results, it appears that the frequency response of the fiber might contribute to the variations observed in the reconstructed sound. Additionally, the reconstruction method involving the mean of each grey pixel value could lead to the rounding of magnitudes for higher frequencies \cite{DEMILIA20132630}. This may result in attenuated amplitudes for outer frequencies in the reconstructed sound. To address this issue, an alternative approach for selecting pixels could be explored, avoiding such attenuation and potentially improving the reconstruction quality of the outer frequency components \cite{Zhu}.

The experiment demonstrates the feasibility of reconstructing external sound signals through a multi-mode fiber. The results of the experiment are analyzed using spectrographic techniques to assess the quality of the reconstructed sound. The use of filtering methods is found to significantly improve the reconstructed sound quality, as evidenced by a comparison of the original and reconstructed signals. The magnitude-squared coherence between the original and reconstructed signals has a remarkable value for frequencies of interest. The visual similarity of the original and the reconstructed sound is apparent in Figure \ref{fig:result}, with the similarity more pronounced and easily perceived through both audio files provided as supplementary materials.

%The visual similarity of the original and the reconstructed sound is apparent in Figure \ref{fig:result}, with the similarity more pronounced and easily perceived through both audio files provided as supplementary materials, accessible at \href{http://ieeexplore.ieee.org}{http://ieeexplore.ieee.org}.

\section*{Conclusion}

In conclusion, our achievement in sound reconstruction from the 128x8 pixel speckle pattern stands as a testament to our success in circumventing the need for additional spatial downsampling. The sensitivity of surface vibrations is notably accentuated in proximity to the fiber source, primarily channeling through a dominant vibrational mode. However, scenarios lacking a singular dominant mode in the multi-mode fiber pose challenges to achieving more precise sound reconstruction. In such intricate scenarios, the employment of signal processing algorithms \cite{Zhu} emerges as a potential strategy to increase the fidelity and audibility of the reconstructed audio.

The demonstrated capabilities of our methodology, with a primary focus on acoustic signal capture, set the stage for further exploration and refinement. This proof-of-concept setup, while showcasing the inherent acoustic sensitivity of fibers, prompts intriguing possibilities for advancement. The method can be expanded in order to exploit the sensitivity of multi-mode fibers to the acoustic signals, for eavesdropping purposes. In addition to that, the deployed methodology offers the potential to be extended for monitoring Earth's crust vibrations and evaluating the structural integrity of civilian buildings, consequently contributing to enhanced safety measures. Hence, the technology ushers in promising prospects for advancing our comprehension of seismic activities and safeguarding vital infrastructure, thereby significantly bolstering urban safety measures and disaster readiness.

\section*{Acknowledgments}
The authors thank Süleyman Kahraman for valuable discussions.

\bibliography{main}

\end{document}